\documentclass[prb,twocolumn,aps,showpacs]{revtex4}
\usepackage{graphics,graphicx,epsfig}
\begin{document} 
\title{Langevin equations for competitive growth models}
\author{F. A. Silveira${}^{1,}$\footnote{Email address: fsilveira@if.uff.br} and
F. D. A. Aar\~ao Reis${}^{2,}$\footnote{Email address: reis@if.uff.br (corresponding author)}}
\affiliation{
Instituto de F\'\i sica, Universidade Federal Fluminense,\\
Avenida Litor\^anea s/n, 24210-340 Niter\'oi RJ, Brazil}
\date{\today}

\begin{abstract}

Langevin equations for several competitive growth models in one dimension are derived.
For models with crossover from random deposition (RD) to some correlated deposition (CD) dynamics,
with small probability $p$ of CD, the surface tension $\nu$ and the nonlinear coefficient
$\lambda$ of the associated equations have  linear dependence on $p$ due 
solely to this random choice. However, they also depend on the regularized
step functions present in the analytical representations of the CD, whose expansion coefficients 
scale with $p$ according to the divergence of local height differences when $p\to 0$. 
The superposition of those scaling factors gives $\nu\sim p^2$
for random deposition with surface relaxation (RDSR) as the CD, and $\nu\sim p$, $\lambda\sim p^{3/2}$
for ballistic deposition (BD) as the CD, in agreement with simulation and other scaling approaches. 
For bidisperse ballistic deposition (BBD), the same scaling of RD-BD model is found.
The Langevin equation for the model with competing RDSR and BD,
with probability $p$ for the latter, is also constructed. It shows linear $p$-dependence of $\lambda$,
while the quadratic dependence observed in previous simulations is explained by an additional crossover
before the asymptotic regime.
The results highlight the relevance of scaling of the coefficients of step function expansions in systems with
steep surfaces, which is responsible for noninteger exponents in some $p$-dependent stochastic equations,
and the importance of the physical correspondence of aggregation rules and equation coefficients.
\end{abstract}

\maketitle

%
\section{Introduction}
\label{intro}

The technological applications of thin films and multilayers motivated intense
theoretical study of growth models in the last decades \cite{barabasi,krug,etb}.
Many processes show evidence of a competition between different aggregation
dynamics. For instance, this occurs in deposition of diamond-like carbon by plasma,
where growth is mainly due to aggregation of slow radicals but ion bombardment
is essential to create $sp^3$ bonds \cite{cn}. When physico-chemical conditions are
continuously changed, such as in cyclical electrodeposition/dissolution of metals,
competing dynamics are also present \cite{shapir}.
Consequently, many competitive growth models were already proposed, with
microscopic aggregation rules representing the atomistic dynamics. They are usually
defined on lattices, such as those
with aggregation of different species of particles \cite{cerdeira,poison1,kotrla} and those
mixing different aggregation rules for the same species
\cite{jullien,albano1,albano2,albano3,tales,chamereis,muraca,kolakowska1,lam,rdcor,albanordcor}.
They usually show crossover effects from one dynamics at small times $t$ and short
length scales $L$ to another dynamics at long $t$ or large $L$, and in special cases
anomalous roughening is present \cite{pradas,companomalous}.

A widely studied group of models is that showing crossover from random (uncorrelated) deposition
(RD) to some correlated deposition (CD) process, hereafter called RD-CD models.
For RD-CD in general, aggregation of each incident particle follows the rules of the CD process
with probability $p$ and those of RD with probability $1-p$ (other models might also 
show the same crossover with a parameter that is not a probability \cite{rdcor}).
Another group of relevant models show crossover from Edwards-Wilkinson (EW) \cite{ew} to
Kardar-Parisi-Zhang (KPZ) \cite{kpz} scaling. A representative model
in this latter group is the competition between ballistic
deposition (BD) \cite{vold} and random deposition with surface relaxation (RDSR, or Family model)
\cite{family}, respectively with probabilities $p$ and $1-p$. In all cases, the crossover appears
for small $p$.

For several reasons, the association of those models with stochastic growth equations of the Langevin type
is a problem of central interest. First,
it facilitates finding asymptotic properties which are frequently unclear in numerical works on
lattice models \cite{predota,chua,haselwandter2006}. Secondly, renormalization study of the growth
equation may relate unexpected numerical results to crossover or instability effects
\cite{haselwandter2008}. 
Finally, improvement in atomistic modeling of thin film growth may be achieved from the advance on
the stochastic equation analysis.

Langevin equations for some of those competitive growth models were
derived in Refs. \protect\cite{muraca,lam,tiago2006} and scaling features of KPZ and EW equations were
discussed by several authors \cite{albano1,rdcor,albanordcor}. For RD-RDSR and BD-RDSR models,
Muraca et al \cite{muraca} suggested quadratic dependence of equation coefficients on the
probability $p$ in the crossover regime, which was in good agreement with available numerical data.
They argued that the time for the less probable process
to occur scales as $1/p$ and that the corresponding increase in local height 
is also proportional to $p$ [Eqs. (1) and (2) of Ref. \protect\cite{muraca}].
The results for RD-RDSR were confirmed by scaling arguments in Refs. \protect\cite{lam,rdcor}.
However, the claim on the universal quadratic form of vanishing coefficients
\cite{muraca} is ruled out by the study of a restricted solid-on-solid (RSOS) \cite{kk} model with
deposition and erosion, which shows linear $p$-dependence of the
nonlinear term in the KPZ equation \cite{tiago2006}, and by the RD-BD model \cite{lam,rdcor},
which shows $p^{3/2}$ scaling of that term.

In this work, Langevin-type equations associated with various competitive lattice
models showing RD-CD and EW-KPZ crossovers are derived through a standard van Kampen expansion 
of the Master equation, followed by a proper choice of the jump moments.
From this approach, the form of the equation coefficients is remarkably
different from the one proposed by Muraca et al \cite{muraca}: in all cases, the random choice of the 
asymptotically dominant process gives a linear dependence on $p$ for the coefficients 
that vanish as $p\to 0$, instead of the quadratic dependence.
However, for RD-CD models, average local slopes diverge as $p\to 0$,
thus the optimal regularization of step functions (present in the transition rates of all
discrete models) have lowest order coefficients that scale as $p$ or $p^{1/2}$, depending
on the aggregation mechanism of the CD. The combination of those scaling relations give
equation coefficients of EW and KPZ equations vanishing as $p$, $p^2$, or $p^{3/2}$. In all cases,
they agree with simulations and other scaling approaches \cite{lam,rdcor}.
For the RDSR-BD model, we show that a non-asymptotic regime with quadratic scaling is present,
which is associated with the dominant effect of subsequent BD events, while the asymptotic linear
relation is predicted for $p$ much smaller than that of previous simulations \cite{chamereis}.

The rest of this work is organized as follows. In Sec. II, three lattice
models with the RD-CD crossover are defined and the approach to predict amplitudes
of roughness scaling is reviewed. In Sec. III, details of the method to derive Langevin equations 
are presented and the equation for the RD-RDSR model is obtained.
In Sec. IV, the equations for the RD-BD model and for the bidisperse ballistic deposition are constructed.
In Sec. V, the equation for the BD-RDSR model is presented and the crossover from
quadratic to linear scaling is discussed.
In Sec. VI we summarize our results and present our conclusions.

%
\section{Lattice models with competition of correlated and uncorrelated deposition}
\label{scalingcompetitive}

In these models, growth begins with a flat $d$-dimensional substrate with cubic symmetry and
$L$ adsorption sites (or columns) in each direction, with a total of $L^d$ sites.
Cubic particles of lateral size $a_\parallel$ (parallel to the substrate plane) and vertical
size $a_\perp$ (parallel to the average growth direction) are sequentially released at
randomly chosen columns above the deposit and fall vertically towards the substrate.
The time interval for deposition of one layer of atoms [$L^d$ atoms] is $\tau$.
Each incident particle may irreversibly stick at the top of the column of incidence,
with probability $1-p$ (RD), or move and stick following some aggregation rule 
that takes into account the neighboring column heights (mimicking physical processes like 
diffusion, desorption, or bond formation), with probability $p$ (CD).

\begin{figure}
\includegraphics[width=7cm]{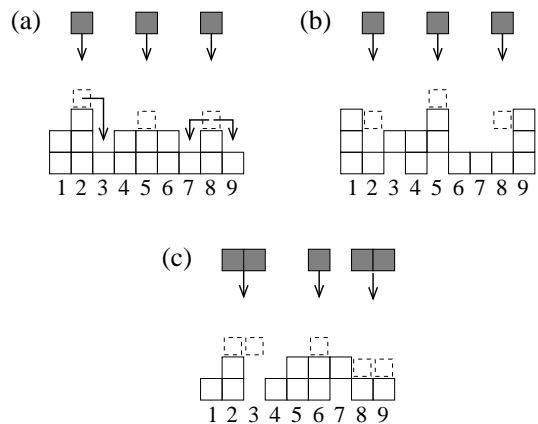}
\caption{Microscopic rules of the models (a) RDSR, (b) BD, and (c) BBD. The incident particles
are shaded squares, the deposited particles are empty squares with solid boundaries and the
squares with dashed boundaries indicate the points of first contact with the surface. In RDSR,
the incident particle may move from the point of first contact, as illustrated by the arrows.
In the other models, aggregation occurs at that point.}
\label{fig1}
\end{figure}

The competitive models where RDSR and BD are the correlated components were introduced by Albano
and co-workers \cite{albano1,albano2}. The rules of RDSR \cite{family} are illustrated in Fig. 1a:
the particle sticks at the top of the
column of incidence if no neighboring column has a smaller height, otherwise it sticks at the top
of the column with the smallest height among the neighbors (if two or more neighbors have the same height,
one of them is randomly chosen). In BD, which is illustrated in Fig. 1b, the incident particle aggregates
at the first site where it finds a nearest neighbor occupied site (lateral or below it) \cite{vold},
which generates a porous deposit.

BBD was introduced in Ref. \protect\cite{bbd1} and is itself defined as a competitive model. 
Taking the $d=1$ case for illustration,
particles of two different sizes incide towards the surface: single-site particles (lateral 
size $a_\parallel$, vertical size $a_\perp$) with probability $1-F$ and double-site particles
(dimers with lateral size $2a_\parallel$, vertical size $a_\perp$) with probability $F$. 
Any incident particle permanently sticks at the first position where it encounters a previously
deposited particle below it. The aggregation of a dimer leads to pore formation when the 
neighboring columns have different heights, similarly to the lateral aggregation of BD.
The rules of BBD are illustrated in Fig. 1c.

In these models, the surface configuration is the array of discrete height variables
${\bf H}=\{h_i\}$, where $i=1,2,3,\dots,L^d$. In all deposits (compact or porous), the height
variable is the one of the highest particle at that substrate
site, thus ${\bf H}$ always refers to the outer surface of the deposit.

The global roughness of the surface is defined as the
rms fluctuation of the height variables, whose average is $\overline{h}$:
\begin{equation}
W(L,t)\equiv \left<  {1\over L^d}\sum_k\left( h_k - \overline{h}\right)^2 \right>^{1/2}.
\label{defw}
\end{equation}
Here the overbars indicate spatial averages over the height variables, the sum is taken over all
the $L^d$ substrate sites, and the angular brackets indicate configurational averages.
At short times, RD dominates, thus the roughness increases as
\begin{equation}
W_{RD}\approx a_\perp{\left( t/\tau\right)}^{1/2} .
\label{wrandom}
\end{equation}
After a crossover time $t_c$, the CD determines the universality class of the process.
The roughness follows Family-Vicsek (FV) scaling \cite{fv} as
\begin{equation}
W(L,t) \approx A L^{\alpha} f\left( \frac{t}{t_\times}\right) ,
\label{fv}
\end{equation}
where $\alpha$ is the roughness exponent, $f$
is a scaling function such that $f\sim 1$ in the regime of roughness saturation
($t\to\infty$) and $t_\times$ is the characteristic time of crossover to
saturation, which scales as 
\begin{equation}
t_\times \approx BL^{z} , 
\label{scalingttimes}
\end{equation}
where $z$ is the dynamic exponent. For $t\ll t_\times$ (but $t\gg t_c$),
the roughness scales as
\begin{equation}
W\approx Ct^{\beta} , 
\label{scalingwgr}
\end{equation}
where $\beta=\alpha/z$ is the growth exponent. In this
growth regime, $f(x)\sim x^{\beta}$ in Eq. (\ref{fv}).

The exponents $\alpha$, $\beta$ and $z$ depend on the basic symmetries of the
CD, but the amplitudes $A$, $B$ and $C$ are model-dependent.
For small $p$, they scale as
   \begin{eqnarray}\label{defdeltaygamma}
    A&\sim&p^{-\delta}                       \nonumber\\
    B&\sim&p^{-y}\\
    C&\sim&p^{-\gamma} ,\nonumber
   \end{eqnarray}
where the convention of crossover exponents $\left( \delta,y,\gamma\right)$ of Albano and co-workers
\cite{albano1} was used.
FV scaling implies
\begin{equation}
y\beta -\delta+\gamma = 0.
\label{scalingalbano}
\end{equation}

The scaling approaches of Refs. \protect\cite{rdcor,albanordcor} explain the values of the exponents
obtained in simulations of lattice models and show that only one exponent [usually $y$ or $\delta$;
Eq. (\ref{defdeltaygamma})] is sufficient to characterize the crossover.

For $p\ll 1$, most deposited atoms attach to the top of the randomly chosen site (RD).
Thus, from Eq. (\ref{wrandom}), the height difference of neighboring sites is of order
${\Delta h}_{RD} \sim a_\perp{\left( \Delta t / \tau\right)}^{1/2}$ after a time interval $\Delta t$.
On the other hand, the average time for a correlated event (probability $p$)
to occur at a given column is $\tau_c\sim\tau/p$.

In the case of ballistic-like models (e. g. BD or BBD), the frequent lateral aggregation
(e. g. at columns 2 and 8 in Fig. 1b) immediately creates correlations
between the neighboring columns. Thus, the time of crossover from random to correlated growth is
\begin{equation}
t_c \sim \tau_c \sim p^{-1}\tau
\label{scalingtc1}
\end{equation}
This applies to other BD-like models, as discussed, e.g., in Ref. \protect\cite{bbdflavio}.

However, in the case of solid-on-solid (SOS) models, such as RDSR, the column height changes by a 
single lattice unit at each time step. A single CD event does not
cancel the random height fluctuation $\Delta h$ of neighboring columns; instead, it only reduces
that fluctuation by one lattice unit (e. g. the particle aggregating at column 3 in Fig. 1a).
The height fluctuation produced by RD will be suppressed
only when the number of correlated events $N_c$ is of order ${\Delta h}_{RD}/a_\perp$.
At the crossover time $t_c$, this number is $N_c = t_c/\tau_c \sim t_c p/\tau$, thus
\begin{equation}
t_c \sim p^{-2}\tau .
\label{scalingtc2}
\end{equation}

In both cases, all time scales of the purely correlated system ($p=1$) are also changed by the scaling
factor $t_c/\tau$, such as the saturation time $t_\times$ [Eq. (\ref{scalingttimes})]. Thus
we have $y=1$ for ballistic-like models and $y=2$ for SOS models with single particle deposition
attempts [Eqs. (\ref{defdeltaygamma})].
This result does not depend on the universality class of the CD neither on the substrate dimension.

The average height difference between neighboring columns saturates at
\begin{equation}
\Delta h \sim a_\perp{\left( t_c / \tau\right)}^{1/2} \sim a_\perp p^{-y/2}
\label{deltahrdcor}
\end{equation}
for all models (ballistic-like and SOS). This is the scaling factor for 
global height fluctuations [Eqs. (\ref{fv}) and (\ref{defdeltaygamma})], thus
\begin{equation}
\delta =y/2 .
\label{deltay}
\end{equation}
Combined with Eq. (\ref{scalingalbano}), it shows that a single exponent ($y$) fully characterizes
the crossover.

%
\section{The RD-RDSR model}
\label{rdrdsr}

The equation associated to the RD-RDSR model is constructed through a 
van Kampen expansion of the Master equation \cite{vankampen},
as discussed in Refs. \cite{haselwandter2006,baggio,vvedensky2003}.

The transition rate $W({\bf H'};{\bf H})$ from the height configuration ${\bf H}=
\{ h_i\}$ to the configuration ${\bf H'}= \{ h_i'\}$ is
   \begin{equation}
   W({\bf H'};{\bf H}) = {1\over \tau} \sum_k{ w_k} \delta \left( h_k', h_k+a_\perp \right)
   \prod_{j\neq k}{ \delta \left( h_j' , h_j\right)} ,
   \label{probtrans}
   \end{equation}
where the $\delta$-function product represents the condition that ${\bf H}$ and ${\bf H'}$ 
differ by the deposition of one only particle, and $w_k$ is the rate at which the process
$h_k\to h_k+a_\perp$ occurs.

Let $K_i^{(1)}$ and $K_{ij}^{(2)}$ be respectively the first and second jump moments of
$W$, computed through \cite{vankampen,moyal}
   \begin{equation}
   K^{(1)}_i({\bf H})=\sum_{\bf H'}(h'_i-h_i)W({\bf H'};{\bf H})
   \label{K1}
   \end{equation}
and
   \begin{equation}
   K^{(2)}_{ij}({\bf H})=\sum_{\bf H'}(h'_i-h_i)(h'_j-h_j)W({\bf H'};{\bf H}).
   \label{K2}
   \end{equation}
According to a theorem of Kurtz \cite{kurtz1,kurtz2}, later revisited 
by Fox and Keiser in the context of a macrovariable description for noisy trajectories 
\cite{foxkeiser} (see also Ref. \cite{nicolis}), we expect that
   \begin{equation}
   {\partial h_i\over\partial t}= K_i^{(1)}({\bf H})+\eta_i(t)
   \label{discretelangevin}
   \end{equation}
gives the macroscopic description of ${\bf H}$ in the hydrodynamic limit. 
If all the conditions imposed in Ref. \cite{vankampen} and in Refs. 
\cite{kurtz1,kurtz2,foxkeiser} are met, the fluctuations in
Eq. (\ref{discretelangevin}) must obey
   \begin{equation}
   \left<\eta_i(t)\right>=0
   \label{m1eta}
   \end{equation}
and
   \begin{equation}
   \left<\eta_i(t)\eta_j(t')\right>=a_\perp K_{i}^{(1)}\delta_{ij}\delta(t'-t),
   \label{m2eta}
   \end{equation}
%
where we used the identity 
\begin{equation}
K_{ij}^{(2)}=a_\perp K_{i}^{(1)}\delta_{ij}
\label{momentrelation}
\end{equation} 
between the first and second jump moments in Eq. (\ref{m2eta}).
In fact, all higher-order jump moments are proportional to $K_{i}^{(1)}$, as can be 
seen from direct calculation.

For the RD-RDSR model (and related competitive models) in $d=1$, the first jump moment 
[Eq. (\ref{K1})] can be cast to the form
   \begin{equation}
    K_i^{(1)}=p{a_\perp\over \tau} \left(\omega_i^{(0)}+\omega_{i+1}^{(1)}+\omega_{i-1}^{(2)}\right)
     +(1-p){a_\perp\over\tau},
   \label{rdsrK}
   \end{equation}
%
where each $\omega_j^{(k)}$ gives the conditions for a particle inciding at column $j$
to move and stick to one of its neighbours or to stick at the incidence column.
Those conditions depend on the local height configuration. 

The variables $\omega_i^{(k)}$ are called {\it aggregation rules}. For the RDSR model, $\omega_i^{(0)}$
represents the conditions for the particle incident at site $i$ to stick at $i$; 
$\omega_i^{(1)}$, the conditions for the particle to relax to its left site ($i-1$);
and $\omega_i^{(2)}$, the conditions for the particle to relax to its right site ($i+1$).
The aggregation rules can be written in terms of discrete step 
functions as \cite{muraca}
   \begin{eqnarray}\label{rdsrrules}
    \omega^{(0)}_i&=& \theta^{i+1}_i\theta^{i-1}_i                        \nonumber\\
    \omega^{(1)}_i&=& {1\over 2}\left(1+\theta^{i+1}_i\right)\left(1-\theta^{i-1}_i\right)\\
    \omega^{(2)}_i&=& {1\over 2}\left(1+\theta^{i-1}_i\right)\left(1-\theta^{i+1}_i\right)\nonumber
   \end{eqnarray}
where $\theta_k^j=\Theta(h_j-h_k)$, and $\Theta(x)$ is the unit step function, defined
at our convenience to be $\Theta(x)=1$ for $x\geq 0$ and $\Theta(x)=0$ for $x<0$.

In order to pass from the discrete model to its continuum limit, we assume there exists 
a continuous function $\Psi(x,t)$ that interpolates all points $h_i(t)$
of the substrate, while $a_\parallel$ is kept small but finite. This is possible if 
we can write
   \begin{equation}
   h_{i\pm n} - h_i = \sum_{k=1}^{\infty}{
   \left( \frac{\partial^k \Psi}{\partial x^k} \right)
   \frac{{\left( \pm a_\parallel n\right)}^k}{k!} } ,
   \label{hdiff}
   \end{equation}
for some $\Psi(x,t)$.

We assume further there also exists an analytical representation of the step 
function $\Theta$ (see for example Refs. \cite{chua,park}), and we define
$\Delta^j_k \equiv h_j-h_k$, so that the function $\Theta(\Delta^j_k)$ can be 
expanded in a power series of the height differences as
%
   \begin{equation}
   \Theta (\Delta^j_k) = 1 + A_{1}\Delta^j_k + A_{2}\left(\Delta^j_k\right)^{2} + \dots
   \label{taylortheta}
   \end{equation}
where the expansion coefficients have to be chosen according to the rules of the lattice
model to be represented \cite{chua,haselwandter2006,haselwandter2008,haselwandter2007}.

Eq. (\ref{rdsrK}) is inserted in Eq. (\ref{discretelangevin}), and
step functions and height differences are expanded according 
to Eqs. (\ref{taylortheta}) and (\ref{hdiff}). Retaining terms up  to the leading order
in $a_\parallel$ and $a_\perp$, and in the limit of small $p$, we obtain
the EW equation \cite{ew}
   \begin{equation}
   {\partial h\over\partial t}= \nu\nabla^2 h+\eta(x,t),
   \label{ew}
   \end{equation}
where
   \begin{equation}
   \nu=\frac{2 a_\perp a_\parallel^2}{\tau}A_1p
   \label{nurdsr}
   \end{equation}
and
   \begin{equation}
   F=\frac{a_\perp}{\tau}.
   \label{Frdsr}
   \end{equation}

These coefficients differ from those of Ref. \protect\cite{muraca}, which gave
$\nu = (2 a_\parallel^2/\tau)A_1 p^2$ and
$F=(a_\perp/\tau)\left[ {\left( 1-p\right)}^2 +p^2\right]$. That work proposes that the height
at a given column increases by a factor proportional to $p$ ($1-p$) after a time interval
$\tau/p$ [$\tau/\left( 1-p\right)$] characteristic of the RDSR (RD) process. This gives the
quadratic dependence of $\nu$ on $p$. However, this hypothesis also leads to a flux $F$
depending on $p$, which is not true. Instead, the model is SOS and no deposition attempt is
rejected, thus the flux is independent of $p$, as given in Eq. (\ref{Frdsr}): one layer of
atoms of height $a_\perp$ is deposited during time $\tau$.

On the other hand, numerical work \cite{albano1} and scaling arguments \cite{lam,rdcor}
give $\nu\sim p^2$ for small $p$, which apparently disagrees with  Eq. (\ref{nurdsr}).
As will be explained below, an additional $p$ factor is hidden in the coefficient $A_1$
of Eq. (\ref{taylortheta}), which is known to be model-dependent.

Since the step function is limited to values $0$ and $1$, the sum of terms in the right-hand side of
Eq. (\ref{taylortheta}) is expected to be of order $1$ or smaller. 
The step at the origin indicates that the first order term $A_{1}\Delta^j_k$ is finite and non-zero.
Indeed, when $A_j$ is computed through some continuous representation of the step function,
as in  Ref. \protect\cite{muraca,vvedensky2003b}, the first order term is of order $1$ for any model.

For pure correlated models, such as RDSR, height differences
$\Delta^j_k$ are of order $1$, thus we sure expect $A_1$ is of that order. On the other hand, in RD-RDSR
with small $p$, typical neighboring height differences $\Delta^j_k$ diverge as $p^{-1}$ 
[Eq. (\ref{deltahrdcor}) with $y=2$]. Thus $A_1$ must vanish as
\begin{equation}
A_1\sim p 
\label{A1rdrdsr}
\end{equation}
for a correct regularization of the step function for small $p$.

Substituting this result in Eq. (\ref{nurdsr}), we obtain $\nu\sim p^2$.
Thus, a first factor $p$ in the surface tension coefficient $\nu$ comes from the random choice
of correlated events, while a second factor $p$ comes
from the reduced smoothing effect of each correlated event in surface steps of depth $p^{-1}$,
as discussed in Sec. II.
This interpretation also differs from that of Muraca et al \cite{muraca}, that relates the
complete $p^2$ factor to the random choice of RDSR. 

Analogous arguments apply to the step function expansion of RD-CD models in general.
From Eq. (\ref{deltahrdcor}), they give
\begin{equation}
A_1\sim p^\delta .
\label{A1rdcd}
\end{equation}
Since the second order term of the expansion in Eq. (\ref{taylortheta}) must also be
of order one, we expect
\begin{equation}
A_2\sim p^{2\delta} .
\label{A2rdcd}
\end{equation}
Eventually the second-order term in Eq. (\ref{taylortheta}) is zero or converges to zero as $p\to 0$,
which would imply that $A_2$ is zero or has a higher power in $p$.

Another important question is a possible crossover effect on the noise term of the stochastic equations.
However, it can be shown that the noise amplitude is independent of the parameter $p$. At the end of
Sec. \ref{rdbdbbd}, a detailed discussion is presented for all models with the RD-CD crossover.

%
\section{The RD-BD and the BBD model}
\label{rdbdbbd}

For the one-dimensional RD-BD model, the first jump moment is
   \begin{eqnarray}
    K_i^{(1)} &=& \frac{p}{\tau} \left[ \omega_i^{(3)}(h_{i-1}-h_i)+\omega_i^{(4)}(h_{i+1}-h_i) +
\omega_{i}^{(5)} a_\perp \right]\nonumber\\
    && +(1-p)\frac{a_\perp}{\tau} ,
   \label{rdbdK}
   \end{eqnarray}
with the aggregation rules given by
   \begin{eqnarray}\label{bdrules}
    \omega^{(3)}_i&=& \theta_i^{i-1}\theta^{i-1}_{i+1}-{1\over 2}\left(1-\theta^i_{i-1}\right)\delta^{i-1}_{i+1}   \nonumber\\
    \omega^{(4)}_i&=& \theta_i^{i+1}\theta^{i+1}_{i-1}-{1\over 2}\left(1-\theta^i_{i+1}\right)\delta^{i+1}_{i-1}  \\
    \omega^{(5)}_i&=& \theta^{i}_{i-1}\theta^{i}_{i+1} , \nonumber 
   \end{eqnarray}
%
where $\delta^{i}_j=\theta^i_j+\theta^j_i-1$ is the Kronecker delta function.
Note that $\omega^{(3)}_i+\omega^{(4)}_i+\omega^{(5)}_i \neq 1$ when $h_i=h_{i-1}$ or $h_i=h_{i+1}$.
However, the corresponding jump moment expression [Eq. (\ref{rdbdK})] is correct, since $\omega^{(3)}_i$ 
and $\omega^{(4)}_i$ are multiplied by $h_{i-1}-h_i$  and $h_{i+1}-h_i$, respectively. Also note
that $\omega^{(3)}_i$ and $\omega^{(4)}_i$ account for lateral aggregation, while $\omega^{(5)}_i$
refers to aggregation at a local maximum (see Fig. 1).

The same steps of the previous model are then followed: Eq. (\ref{rdbdK}) is inserted in Eq.
(\ref{discretelangevin}), with step functions and height differences expanded according 
to Eqs. (\ref{taylortheta}) and (\ref{hdiff}). Retaining terms up to the leading order in 
$a_\parallel$ and $a_\perp$, and in the small $p$ limit, we obtain the KPZ equation
   \begin{equation}
   \frac{\partial h}{\partial t}= F + \nu\nabla^2 h+\frac{\lambda}{2} {\left(\nabla h\right)}^2 +\eta(x,t) 
   \label{kpz}
   \end{equation}
with
   \begin{equation}
   \nu = {a_\parallel^2\over\tau} p ,
   \label{nurdbd}
   \end{equation}
   \begin{equation}
   \lambda = {10 a_\parallel^2\over\tau}A_1p ,
   \label{lambdardbd}
   \end{equation}
and
   \begin{equation}
   F = \frac{a_\perp}{\tau} .
   \label{Frdbd}
   \end{equation}

A naive inspection of Eqs. (\ref{nurdbd}) and (\ref{lambdardbd}) suggests $\nu\sim p$ and
$\lambda\sim p$. However, the scaling of $A_1$ and $A_2$ in Eqs. (\ref{A1rdcd})
and (\ref{A2rdcd}), with $y=1$ for ballistic-like models, gives
\begin{equation}
\nu\sim p , \lambda\sim p^{3/2} ,
\label{nulambdaRDBD}
\end{equation}
which agrees with simulation \cite{albano2} and scaling arguments \cite{lam,rdcor}.

Thus, this model also shows that the scaling of the equation coefficients on $p$ depend not
only on the random choice of BD but also on the scaling of height differences.
Moreover, our analysis show that equation coefficients which are noninteger powers or $p$, such as
$p^{3/2}$, can be predicted by construction of growth equations from the microscopic rules,
with the noninteger exponent related to the step function regularization.

The coefficient of the surface tension term [Eq. (\ref{nurdbd})] is always positive.
Indeed, the lateral aggregation rules of BD [$\omega^{(3)}_i$ and $\omega^{(4)}_i$ 
in Eq. (\ref{bdrules})] reduce local height differences, as illustrated by deposition
at columns $2$ and $8$ in Fig. 1b, which is the role of surface tension.
This balances the negative contribution to the surface tension from
aggregation at a local maximum, illustrated by deposition at column $5$ in Fig. 1b. 

This result differs from what is obtained with the aggregation rules of Ref. \protect\cite{muraca}: 
$\nu = -({a_\parallel^2a_\perp/\tau})A_1 p$, which is negative for (expected) positive $A_1$.
With the aggregation rules presented in that work, lateral aggregation does not contribute
to surface tension, thus aggregation at a local maximum renders $\nu$ negative, as can be inferred
from the $a_\perp$ factor appearing in that formula for $\nu$ (in Ref. \cite{muraca}, it follows only 
from the expression of $\omega^{(5)}_i$).

The preceding discussion shows that
it is essential to check the consistency of the equation coefficients and the geometry of
the model when strong approximations (such as regularization of step functions) are involved.
When the surface is rough, pure lateral aggregation
at edges and steps tends to bring the surface to a plain surface state, through both 
non-conservative (taming of a height difference larger than $a_\perp$) and conservative
mechanisms (taming of a step of height $a_\perp$). Indeed,
we chose aggregation rules for BD such that the pure lateral aggregation gives a positive
contribution to the laplacian term, as it is expected. 
That was only possible by allowing
$\omega^{(3)}_i+\omega^{(4)}_i+\omega^{(5)}_i \neq 1$
in Eq. (\ref{bdrules}) and avoiding products between terms such as $1-\theta^i_j$,
which will be subsequently regularized.

Now we consider the BBD model. The first jump moment is
   \begin{eqnarray}
    K_i^{(1)} &=& \frac{p}{\tau} [ \left(\omega_i^{(6)}+\omega_{i-1}^{(9)}\right) \left( a_\perp+h_{i-1}-h_i\right)\nonumber\\
                 &&+ \left( \omega_i^{(7)}+\omega_{i+1}^{(8)}\right) \left(a_\perp+h_{i+1}-h_i\right) ]\nonumber\\
              &&\left. + a_\perp\omega_{i+1}^{(6)}+a_\perp\omega_{i-1}^{(7)}+a_\perp\omega_i^{(9)}+
                   a_\perp\omega_i^{(8)} \right] \\
              &&+ \left( 1-p\right) \frac{a_\perp}{\tau} ,\nonumber
   \label{bbdK}
   \end{eqnarray}
with
   \begin{eqnarray}\label{bbdrules}
    \omega^{(6)}_i &=&  {(1/2)}\theta^{i-1}_i     \nonumber\\
    \omega^{(7)}_i &=&  {(1/2)}\theta^{i+1}_i   \\
    \omega^{(8)}_i &=&  {(1/2)}\left(1-\theta^{i-1}_i\right) \nonumber\\
    \omega^{(9)}_i &=&  {(1/2)}\left(1-\theta^{i+1}_i\right) , \nonumber
   \end{eqnarray}
where the factors $1/2$ correspond to the equal probability for the two possible 
orientations of the dimers on the incidence site.
The KPZ equation [Eq. (\ref{kpz})] is also obtained from this rules. In the small 
$p$ limit, it has coefficients
   \begin{equation}
   \nu = {a_\parallel^2\over\tau} p ,
   \label{nubbd}
   \end{equation}
   \begin{equation}
   \lambda = 4\frac{a_\parallel^2}{\tau} A_1 p ,
   \label{lambdabbd}
   \end{equation}
and
   \begin{equation}
   F = \frac{a_\perp}{\tau}\left( 1+p\right) .
   \label{Fbbd}
   \end{equation}
The $p$-dependence of $A_1$ for ballistic-like models [Eq. (\ref{A1rdcd}) with $\delta=1/2$]
again gives $\nu\sim p$ and
$\lambda\sim p^{3/2}$, in agreement with Ref. \protect\cite{bbdflavio}, which combined scaling
properties of the KPZ equation in one dimension and numerical results.

The relation between the first and second jump moments (Eq. \ref{momentrelation})
shows that this scaling picture can also provide the noise term of the growth equation.
We follow Ref. \cite{barabasi} and rewrite Eq. (\ref{m2eta}) as 
$\left<\eta_i(t)\eta_j(t')\right>=D\delta_{ij}\delta(t'-t)$ to have $D= a_\perp K_i^{(1)}$,
where $D$ is the amplitude of the noise correlations.
For each model, we expand the step functions and height differences
(Eqs. \ref{taylortheta} and \ref{hdiff})
and retain terms up to the leading order in $a_\perp$ and $a_\parallel$. This gives
$D\sim a_\perp F$ for all models, as can be found from inspection of Eqs. (\ref{nurdsr}) and (\ref{Frdsr}),
(\ref{nurdbd}) to (\ref{Frdbd}), (\ref{nubbd}) to (\ref{Fbbd}) and (\ref{nurdsrbd}) to (\ref{Frdsrbd}),
since all time and space derivatives of $h$ are finite in the limiting process.
That means $D=a_\perp^2/\tau$ plus terms of order $a_\perp a_\parallel^2$ or $a_\perp^2a_\parallel^2$
in all RD-CD models. For BBD, $D=a_\perp^2(1+p)/\tau +{\cal O}(a_\perp a_\parallel^2)$,
due to the particular choice of the time unit for that model.

These results show that, in the
small $p$ limit, there is no effect of this parameter on the noise amplitude. Consequently, all the
crossover effects depend on the coefficients $\nu$ and $\lambda$ (in contrast to what is observed
in other growth models \cite{haselwandter2010}).

%
\section{The RDSR-BD model}
\label{rdsrbd}

This model was introduced in Ref. \protect\cite{jullien} and involves the competition of BD (KPZ class),
with probability $p$, and RDSR  (EW class), with probability $1-p$. In Ref. \protect\cite{chamereis},
scaling properties were studied numerically, with the coefficient of the nonlinear term
scaling as $\lambda\sim p^2$ for $0.2\leq p\leq 0.5$.
That quadratic dependence was proposed analytically by Muraca et al \cite{muraca}.

The first jump moment in this case is
\begin{equation}
K_i^{(1)} =  p{\cal K}_i^{BD}+(1-p){\cal K}_i^{RDSR}
\label{rdsrbdK}
\end{equation}
where ${\cal K}_i^{RDSR},{\cal K}_i^{BD}$ are the first jump moments
   \begin{equation}
    {\cal K}_i^{RDSR}={a_\perp\over \tau} \left(\omega_i^{(0)}+\omega_{i+1}^{(1)}+\omega_{i-1}^{(2)}\right)
   \label{Krdsr}
   \end{equation}
and
   \begin{equation}
    {\cal K}_i^{BD} = {1\over \tau} \left[ \omega_i^{(3)}(h_{i-1}-h_i)+\omega_i^{(4)}(h_{i+1}-h_i) +
\omega_{i}^{(5)} a_\perp \right] ,
   \label{Kbd}
   \end{equation}
where $\omega_i^{(k)}$, $k=0,\dots 5$,
are the  aggregation rules given in Eqs. (\ref{rdsrrules}) and (\ref{bdrules}).
Following the same approach of the other models, we obtain the KPZ equation with coefficients
   \begin{equation}
   \nu = {a_\parallel^2\over\tau}p+ {2 a_\parallel^2a_\perp\over\tau} A_1\left(1-{3\over 2}p\right),
   \label{nurdsrbd}
   \end{equation}
   \begin{equation}
   \lambda = \frac{2 a_\parallel^2}{\tau}A_1\left(5-a_\perp A_1\right)p ,
   \label{lambdardsrbd}
   \end{equation}
and
   \begin{equation}
   F = \frac{a_\perp}{\tau} .
   \label{Frdsrbd}
   \end{equation}
Both RDSR and BD have correlated kinetics which lead to finite average values of local slopes,
even in the steady states. Consequently, the leading coefficients of the step functions ($A_1$, $A_2$)
do not vanish in the limit $p\to 0$, in contrast with the models with crossover from RD.
Thus, Eq. (\ref{lambdardsrbd}) gives $\lambda \sim p$ as $p\to 0$, while the other coefficients
remain nonzero.

This result disagrees with the quadratic dependence observed in simulations of Ref. \protect\cite{chamereis}
and suggested in analytical work of Ref. \protect\cite{muraca}.
In order to understand this discrepancy, the conditions where the BD component generates nonlinearity
in the RDSR-BD model have to be analyzed.

In Fig. 1b, deposition at columns $2$ and $8$ shows the condition in which lateral aggregation
(characteristic of BD) leads to excess velocity:
deposition occurs at a column $i$ which has at least one neighbor with height larger by $2a_\perp$ or more.
This leads to formation of a hole in column $i$. However, for small $p$, pure RDSR dominates.
%
We simulated the one-dimensional RDRS model in lattice sizes $L=256$ and $L=512$ and found that
the number of columns where excess velocity is possible is $P\approx 0.044$.
This fraction is small because RDSR produces a very smooth surface, with a very small number of high steps.
On the other hand, in pure BD, our simulations show that this probabillity is near $1/2$.
Thus, in the competitive model with small $p$, the fraction of columns which have lateral growth
(i. e. nonlinear growth) is approximately $Pp$.

On the other hand, the BD model itself creates conditions for two neighboring sites to have height
difference $2a_\perp$ or more: if a column $i$ has one larger neighbor $j$ ($h_i<h_j$), a BD event at $j$
followed by another BD event at $i$ leads to lateral aggregation with formation of a hole.
For instance, this would correspond to the deposition in column $j=5$, shown in Fig. 1b, followed by
deposition in column $i=4$ (not shown). 
In the pure RDSR surface, the fraction of columns with at least
one higher neighbor is $Q\approx 0.44$, also obtained from simulation. Thus, the probability of this column
having nonlinear growth due to those subsequent BD events is approximately $Qp^2$.

From the point of view of Eq. (\ref{lambdardsrbd}), pure RDSR corresponds to $A_1\sim P$ in
the regularization of step functions, while a BD event corresponds to $A_1\sim Qp$. 

For small enough $p$, we certainly have $Pp>Qp^2$, thus the crossover EW-KPZ is dominated by
BD events taking place on a nearly pure RDSR surface. This occurs for $p<P/Q\approx 0.10$.
In this regime, the linear dependence of $\lambda$ on $p$ [Eq. (\ref{lambdardsrbd})] is expected.
However, simulation results of Ref.
\protect\cite{chamereis} are for $p\geq 0.2$. In the lower limit $p=0.2$, we have
$Pp\approx 0.0088$ and $Qp^2\approx 0.0176$. This means that $Qp^2$ is twice as large as
$Pp$, and their difference is enhanced for larger $p$. 
Consequently, the simulated range
of $p$ favors nonlinearities arising from two subsequent BD events at neighboring columns, which
explains the observed quadratic dependence of the coefficient $\lambda$ on $p$.

The arguments of Ref. \protect\cite{muraca} for the $\lambda\sim p^2$ behavior were also based on
the association of the $p^2$ factor to the random choice of BD. However, it is also a double counting
of the factor $p$, which is reasonable for the simulated range of $p$ but fails for very small $p$.

The linear scaling of the coefficient $\lambda$ on $p$ was also
found in the RSOS model of deposition and erosion of Ref. \protect\cite{tiago2006}, both in simulations
and in the derivation of the associated KPZ equation. The constraint on the neighboring height difference
of the RSOS model leads to rejection of deposition and erosion attempts, which is the mechanism to generate
nonlinear growth. That rejection occurs with a probability much larger than the probability $P$ for the RDSR,
thus the linear dependence on $p$ was easily observed in simulations \cite{tiago2006}.

%
\section{Conclusion}

Langevin equations associated with various competitive lattice models were derived.
The approach is based on a van Kampen expansion of the Master equation,
but the correct assessment of how do characteristic times and lengths scale with
the competing parameter
plays a central role if we are to find the true dependence of the
equation coefficients in the crossover regimes.
Moreover, it is essential to choose representations of aggregation rules 
(using e. g. step and delta functions) that lead to physically
reasonable equation coefficients, as RDSR and BD models illustrate.

We considered a series of models with crossover from random deposition to
correlated growth (RD-CD), with probability $p$ for the latter,
and a model with EW-KPZ crossover, with probability $p$ for the KPZ component.
All coefficients that vanish as $p\to 0$ show a linear $p$ dependence
arising from the random choice of aggregation rules. However, in the RD-CD case,
neighboring height differences diverge in that limit, which leads to the $p$-scaling of
the parameters of the optimal regularization of step functions. Thus, the
coefficients depending on those parameters show scaling as $p$, $p^2$, and $p^{3/2}$,
in all cases in agreement with simulation results and other scaling approaches.
For the model with EW-KPZ crossover, the quadratic dependence of the nonlinear term coefficient,
observed in simulations, is explained as a crossover behavior due to particular model
features, while linear $p$ dependence is expected for very small $p$.
Although the scaling properties derived here are similar to previous works on those models
\cite{muraca,lam,rdcor}, the interpretation is very different and the applicability of
the method is broader, for instance being extendable to higher dimensions.

\acknowledgments

This work was partially supported by CNPq and FAPERJ (Brazilian agencies).


\vfill\eject

\end{document}